\documentclass[useAM,usenatbib]{mn2e}
\usepackage{epsfig, bm}

\def\s{{\rm\,s}}
\def\erg{{\rm\,erg}}
\def\ev{{\rm\,eV}}
\def\eV{{\rm\,eV}}
\def\ang{{\rm\,\AA}}
\def\cm{{\rm\,cm}}

\def\mum{{\rm\,\mu m}}
\def\km{{\rm\,km}}

\def\kev{{\rm\,keV}}

\def\K{{\rm\,K}}
\def\yr{{\rm\,yr}}
\def\rt{R_{\rm T}}
\def\rs{R_{\rm S}}

\def\mbh{M_{\rm BH}}
\def\msun{M_\odot}

\def\rstar{R_\star}
\def\rdisk{R_{\rm disk}}
\def\rp{R_{\rm p}}

\def\rl{R_{\rm L}}
\def\rphes{R_{\rm ph,es}}
\def\tphes{T_{\rm ph,es}}
\def\redge{R_{\rm edge}}
\def\rsource{R_{\rm source}}
\def\vwind{v_{\rm wind}}

\def\vlos{v_{\rm LOS}}
\def\tfall{t_{\rm fallback}}
\def\tedd{t_{\rm Edd}}
\def\tlte{t_{\rm LTE}}

\def\trattenday{\left(\frac{t}{\rm 10\,day}\right)}
\def\mdotedd{\dot{M}_{\rm Edd}}
\def\mdotfb{\dot{M}_{\rm fallback}}
\def\mdotrat{\left(\frac{\dot{M}_{\rm fallback}}{\dot{M}_{\rm Edd}}\right)}
\def\mbhrat{M_6}
\def\rprat{R_{\rm p, 3\rs}}
\def\rprtrat{\left(\frac{\rp}{\rt}\right)}
\def\mstarrat{m_\star}
\def\rstarrat{r_\star}
\def\lesssim{\mathrel{\hbox{\rlap{\hbox{\lower3.5pt\hbox{$\sim$}}}\hbox{$<$}}}}
\def\gtrsim{\mathrel{\hbox{\rlap{\hbox{\lower3.5pt\hbox{$\sim$}}}\hbox{$>$}}}}

\title[Spectroscopic Signatures of the Tidal Disruption of Stars]{Spectroscopic Signatures of the Tidal Disruption of Stars by Massive Black Holes}
\author[L. E. Strubbe and E. Quataert]{Linda E. Strubbe$^{1}$\thanks{E-mail:
linda@astro.berkeley.edu}, Eliot Quataert$^{1}$ \\ 
$^{1}$Astronomy Department and Theoretical Astrophysics Center, 
601 Campbell Hall, University of California, Berkeley CA, 94720, USA\\}

\begin{document}
\date{Accepted . Received ; in original form }

\pagerange{\pageref{firstpage}--\pageref{lastpage}} \pubyear{2011}
\maketitle

\label{firstpage}
\begin{abstract}
  During the tidal disruption of a main sequence star by a massive
  black hole (BH) having mass $\mbh \lesssim 10^7 M_\odot$, the
    stellar debris is expected to fall back to the BH at a rate well
    above the Eddington rate.  Some fraction of this gas is predicted
    to be blown away from the BH, producing an optically bright flare
    of radiation.  We predict the spectra and spectral evolution of
  tidal disruption events, focusing on the signatures produced by
  photoionized gas outside the photosphere of this super-Eddington
  outflow.  We show that the spectrum of such an outflow should
    show absorption lines that are strongly blueshifted relative to
  the host galaxy, are typically very broad ($0.01-0.1c$), and are
  most prominent at ultraviolet wavelengths (e.g., C IV, Lyman
  $\alpha$, O VI) at early times ($\lesssim \, 1$ month for a $\sim
  10^6$ $\msun$ BH).  There may also be optical absorption lines of
  hydrogen and He II if there is a lower velocity component to the
  outflow ($\lesssim 0.01$ c).  At later times, the outflow falls out
  of thermal equilibrium and the continuum emission likely
  hardens---the absorption lines will then primarily be in the soft
  X-rays.

  Supernovae in galactic nuclei are a significant source of confusion
  in optical surveys for tidal disruption events: we estimate that
  nuclear Type Ia supernovae are two orders of magnitude more common
  than tidal disruption events at $z \sim 0.1$ for ground-based
  surveys.  Nuclear Type II supernovae occur at a comparable rate but
  can be excluded by pre-selecting red galaxies.  The contamination
  from nuclear supernovae can be reduced to a manageable level by
  using high-resolution follow-up imaging with adaptive optics or the
  Hubble Space Telescope.  Our predictions should help optical
  transient surveys capitalize on their potential for discovering
  tidal disruption events.

\end{abstract}


\section{Introduction}\label{intro}

A massive black hole (BH) at the center of a galaxy can tidally
disrupt stars that pass within a radius $\rt \sim \rstar(\mbh/M_{\rm
  star})^{1/3}$ of it, where $\mbh$ and $M_{\rm star}$ are the mass of
the BH and star\footnote{We reserve the symbol $M_\ast$ for the
  stellar mass of a galaxy (\S \ref{SNrates}).}, respectively, and
$\rstar$ is the radius of the star.  For solar-mass stars, $\rt$ lies
outside the BH's event horizon for $\mbh \lesssim 10^8\msun$; in these
systems, a fraction of the stellar debris is expected to flow back
towards the BH following disruption, releasing a flare of radiation.
Although the rate of tidal disruption events per galaxy is uncertain,
it is likely $\sim 10^{-6} - 10^{-3}$ per year \citep[e.g.,][]{mt99,
  wm04, donley}.

Observing and studying such flares has the potential to inform our
understanding of accretion physics, the mass function of BHs, and the
dynamics of stars in the nuclei of galaxies.  To date, several
candidate events have been discovered, and many more are likely to be
found in the coming years.  A handful of candidates were discovered in
soft X-rays by the ROSAT All-Sky Survey and XMM-Newton Slew Survey,
and several more candidates have been discovered in the ultraviolet
(UV) by GALEX \citep{komossa, gezari08, gezari09}; these observations
likely probe emission from an accretion disk close to the BH.
Searches with GALEX are ongoing, and several new wide-field,
high-cadence optical transient surveys have recently started or are
planned: the Palomar Transient Factory (PTF), Pan-STARRS, and the
Large Synoptic Survey Telescope (LSST).

In \citet{sq09}, we argued that optical surveys have the potential to
discover tens to hundreds of tidal disruption events per year.  Bright
optical emission occurs when stellar debris is unbound by the intense
radiation pressure produced by the debris falling back to the BH at a
super-Eddington rate.  We predicted that this outflow produces a
  flare as bright as a supernova, $\sim 10^{43}-{\rm
    few}\times10^{44}\erg\s^{-1}$, that lasts for days to weeks.
These optically luminous flares may be detectable out to $z \sim 1$
(and perhaps beyond).  However, optical transient surveys are finding
and will find {\it many} bright transients close to the centers of
galaxies (e.g., active galactic nuclei and supernovae).  Identifying
tidal disruption events amid an array of far more common transient
phenomena thus poses a substantial challenge.  Observational follow-up
is crucial: detailed multi-wavelength light curves, high-resolution
imaging (to determine that events are truly nuclear), and spectroscopy
are all required.  In the long term, the last of these has the
potential to be the most definitive signature of a tidal disruption
event.

In this paper, we predict the optical--X-ray spectroscopic signatures
of tidal disruption events as a function of time, focusing on the
outflows produced when the fallback rate is super-Eddington.  In such
outflows, gas outside the electron scattering photosphere emits
photons and absorbs photons from deeper in, producing a spectrum that
can contain emission and absorption features.
A separate source of spectroscopic features can arise at late times,
after the outflow subsides, from the half of the star that gained
energy upon disruption and is escaping from the BH in the star's original
orbital plane.  The surface of that equatorial material is irradiated
by the accretion disk, producing broad emission lines (mostly
hydrogen) offset in velocity from the galactic lines \citep{sq09}.
Detecting these lines would also be a strong confirmation of a tidal
disruption event, but they are usually substantially fainter than the
spectral diagnostics presented here and so will be more difficult to
observe.

The remainder of the paper is organized as follows: in
\S\ref{outflowprops}, we review the physics of super-Eddington
outflows produced during tidal disruption events and describe how we
use the photoionization code Cloudy to calculate the spectral lines of
the outflow and how we calculate the line profiles. We also critically
assess when the assumption of thermal equilibrium for the outflow's
emission employed by \citet{sq09} is valid (\S \ref{sec:te}).  In
\S\ref{results} we describe our primary spectroscopic predictions.  We
then briefly estimate the rate of supernovae in the nuclei of galaxies
(\S \ref{SNrates}), since nuclear supernovae are one of the primary
sources of confusion in optical searches for tidal disruption
events. Finally, in \S\ref{discussion} we discuss our results and
their implications for observing and identifying tidal disruption
events.

\vspace{-0.4cm}
\section{Super-Eddington Outflows}\label{outflowprops}
\vspace{-0.1cm}
\subsection{Summary of Basic Properties}\label{basicprops}

We summarize theoretical expectations for the physics of tidal
disruption events.  Following a star's disruption, roughly half of the
stellar debris becomes bound to the BH, falls back to pericenter, and
shocks; the rate of fallback is \citep{rees,phinney89}
\begin{equation}
\dot{M}_{\rm fallback} \simeq \frac{1}{3}\frac{M_{\rm star}}{t_{\rm fallback}}\left(\frac{t}{t_{\rm fallback}}\right)^{-5/3}
\label{mdotfallback}
\end{equation}
where 
\begin{equation}\label{tfallback}
t_{\rm fallback} \simeq 20 \mbhrat^{5/2}\rprat^3 \rstarrat^{-3/2} {\, \rm min}
\nonumber
\end{equation}
is the period of the most bound debris, the BH mass is
$\mbh\equiv\mbhrat \times 10^6\msun$, the pericenter distance of the
star's orbit is $\rp$, $\rprat\equiv\rp/3\rs$ (where $\rs$ is the
Schwarzschild radius), and the stellar radius $\rstarrat\equiv
R_\star/R_\odot$.  For $\mbh \lesssim {\rm few} \times 10^7 M_\odot$,
the fallback rate predicted by equation (\ref{mdotfallback}) can be
much greater than the Eddington rate $\dot{M}_{\rm Edd}$ for a period
of weeks to years; here $\dot{M}_{\rm Edd} \equiv 10L_{\rm Edd}/c^2$,
$L_{\rm Edd}$ is the Eddington luminosity, and 0.1 is the fiducial
efficiency of converting accretion power to luminosity.

While the fallback rate is super-Eddington, the stellar gas returning
to pericenter is so dense that the photons produced in the shock are
unable to escape and cool the gas; in particular, the time for photons
to diffuse out of the gas is longer than both the inflow time in the
disk and the dynamical time characteristic of an outflow.  The gas is
likely to form an advective accretion disk accompanied by powerful
outflows \citep[e.g.,][]{king03,ohsuga05}.

In \citet{sq09}, we developed a simple model to describe the
outflowing gas \citep[see also related estimates in][]{king03,rossi09}.
We assume the outflowing gas is launched from $\sim \rl \equiv 2\rp$
at a rate
\begin{equation}
 \dot{M}_{\rm out} \equiv f_{\rm out}\mdotfb 
\label{mdotout}
\end{equation}
with a terminal velocity
\begin{equation}
  \vwind \equiv f_v v_{\rm esc}(\rl),
\label{vwind}
\end{equation}
which is typically $1-10\%$ of the speed of light.  Radiation
hydrodynamical simulations of super-Eddington accretion show that the
density and velocity structure of the outflowing gas varies with
angle, with higher speed outflows along the pole
\citep[e.g.,][]{ohsuga05}; in the tidal disruption context, the outflow
properties may also vary with time.  To account for variations with
viewing angle, we consider values of $f_{\rm out}$ ranging from 0.01
to 0.3, with $f_{\rm out} = 0.1$ as our fiducial value, and values of
$f_v$ ranging from 0.1 to 1, with $f_v=1$ as our fiducial value.  We
approximate the outflow's geometry as spherical, with a density
profile
\begin{equation}\label{densityprofile}
\rho(r,t) \simeq \frac{\dot{M}_{\rm out}(t-r/\vwind)}{4\pi r^2 \vwind} 
\end{equation}
inside the outflow where $r \lesssim \redge \equiv \vwind t$.  For $r
\ll \redge$, the density varies as $\rho(r,t) \sim \dot{M}_{\rm
  out}(t)/4\pi r^2 \vwind$.  When $t \gtrsim {\rm few}\times \tfall$,
the density increases with radius approaching $\redge$: most of the
mass is near the edge, within a shell of thickness $\Delta r_{\rm
  shell}$, so that
\begin{equation}\label{shell_rho}
\rho(\redge,t) \sim 
 \frac{\frac{1}{2}f_{\rm out}M_{\rm star}}{4\pi \redge^2 \Delta r_{\rm shell}} \, .
\end{equation}
Since most of the gas was expelled during a period lasting
$\sim\tfall$, $\Delta r_{\rm shell}$ is at least $\sim \vwind\tfall$;
to account for a possible variation in outflow velocity during the
period when most of the wind is launched, we assume that the wind
speed varies by $\Delta \vwind/\vwind \sim 10 \%$, so that $\Delta
r_{\rm shell} \sim \max(\vwind \tfall,0.1\vwind t)$.  The exact
magnitude of $\Delta \vwind$ is uncertain, but accounting for this
systematic variation is important because otherwise the shell is
unphysically narrow and dense at late times.

At most wavelengths, the dominant opacity in the outflow is electron
scattering.  The outflow is optically thick to electron scattering out
to a radius $\rphes$, the electron scattering photosphere, at which
$\rphes\rho(\rphes)\kappa_{\rm es} \sim 1$, where $\kappa_{\rm es}$
the opacity to electron scattering:
\begin{equation}
\rphes \simeq 10 f_{\rm out}f_v^{-1}\mdotrat \rprat^{1/2}\rs \, .
\label{rphes}
\end{equation}

Because they are trapped by electron scattering, photons produced in
the shock cool adiabatically as the gas expands in the outflow.  In
\citet{sq09} we assumed that the gas and photons would be in thermal
equilibrium at the shock so that the outflowing photons would have a
blackbody spectrum, given by
\begin{equation}\label{Loutflow}
\nu L_\nu = 4\pi^2 \rphes^2 \nu B_\nu(\tphes) \, 
\end{equation}
where $\tphes$ is the temperature at the electron scattering
photosphere.  The flow remains supported by radiation pressure and
thus adiabatic expansion causes the temperature in the outflow to
scale as $T \propto \rho^{1/3}$.  In thermal equilibrium, the
temperature of the gas and radiation at the shock ($T_{\rm eq}$) are
determined by $aT_{\rm eq}^4 \simeq u_{\rm pre,gas}$, i.e., the
post-shock photon energy density is approximately equal to the bulk
kinetic energy density of the pre-shock gas.  We approximate that the
gas falls back to pericenter spherically, so that $u_{\rm pre,gas}
\sim \mdotfb v_{\rm esc,L}/4\pi \rl^2$, where $v_{\rm esc,L}$ is the
escape velocity of the gas at $\rl$.  The temperature at $\rphes$ in
thermal equilibrium is thus
\begin{equation}\label{tphnoedge}
\tphes \sim  2 \times 10^5 \, {\rm K} \left(\frac{f_v}{f_{\rm out}}\right)^{1/3}
\!\!\mdotrat^{-5/12}\!\!\!\!\mbhrat^{-1/4}\rprat^{-7/24}.
\end{equation}

After weeks to months, the outflow finally becomes optically thin to
electron scattering, revealing the accretion disk close to the BH.
The accretion disk's spectrum is a multicolor blackbody with
temperatures $\sim 10^5\K$, described in more detail in \citet{sq09}.
Eventually, after a time
\begin{equation}
  t_{\rm Edd} \simeq 0.1 \, \mbhrat^{2/5}\rprat^{6/5}\mstarrat^{3/5}
\rstarrat^{-3/5}\yr
\label{tEdd}
\end{equation}
(where $\mstarrat\equiv M_{\rm star}/\msun$),
the mass fallback rate decreases below the Eddington rate, and
radiation pressure is no longer strong enough to unbind new gas.  The
previously expelled gas continues to expand outwards, becoming a thin
shell located at $\sim \redge$ with a thickness $\Delta r_{\rm shell}$
and a density given in equation (\ref{shell_rho}).  From some viewing
angles, the accretion disk will continue to be seen through this
shell, with the exact covering fraction of the shell depending on the
geometry of the outflow at early times, which is somewhat uncertain.

\vspace{-0.4cm}
\subsection{The Applicability of Thermal Equilibrium}
\label{sec:te}

Equations (\ref{Loutflow}) and (\ref{tphnoedge}) assume that the gas and
radiation are thermally well-coupled from the shock at $\sim \rp$ to
the outflow's electron scattering photosphere at $\rphes$.  However,
the post-shock gas and radiation may not have time to reach thermal
equilibrium before advecting away from the shock.  Here we
quantitatively assess the applicability of thermal equilibrium drawing
an analogy to the radiation-mediated shocks present during supernovae
and shock-breakout (e.g., \citealt{katz10}).

In local thermal equilibrium (LTE), the temperature of the gas and
radiation are given by $T_{\rm eq}$, where $aT_{\rm eq}^4 \simeq
u_{\rm pre,gas}$ (see the text before eq. [\ref{tphnoedge}]).  The
dominant continuum emission process is free-free emission.  
The time to reach thermal equilibrium $\tlte$ is determined by the
timescale for free-free emission in the post-shock plasma to produce
the number density of photons required for thermal equilibrium.  Prior
to $\sim t_{\rm LTE}$, the gas and radiation are instead in Compton
equilibrium at a temperature $T_{\rm shock}$ that is substantially
larger than $T_{\rm eq}$ \citep{katz10}.
The shock jump conditions imply that the velocity falls and the
density of gas rises by a factor of 7 as the gas moves from pre-shock
to post-shock.  Using equation (13) from \citet{katz10}, we then find
that the time for the post-shock plasma to thermalize, in units of the
local dynamical time $t_{\rm dyn} = \rl/v_{\rm esc,L}$, is given
by\footnote{This estimate of $\tlte$ is about 100 times shorter
  than the most na{\"i}ve estimate of the thermalization time, $\sim
  (\alpha_{\rm ff}^Rc)^{-1}$, where $\alpha_{\rm ff}^R$ is the
  Rosseland mean absorption coefficient for free-free interactions.
  The time $t_{\rm LTE}$ is that associated with waiting for the gas
  to emit photons rather than waiting for it to absorb photons already
  present. The large numerical difference between $t_{\rm LTE}$ and
  $\sim \alpha_{\rm ff}^R c$ is primarily due to different weighting
  of frequencies in the averaging; in the correct calculation,
  emission of low-frequency photons dominates \citep{katz10}, while in
  $\alpha_{\rm ff}^R$, absorption of photons close to the blackbody
  peak dominates.} 
\begin{eqnarray} \label{therm}
\frac{\tlte}{t_{\rm dyn}} &\sim& 400 \, \mbhrat^{-5/8}\rprat^{-47/16}\left(\frac{\rstarrat}{\mstarrat}\right)^{9/8}\trattenday^{15/8} \nonumber \\
& \sim & 0.9 \, \mbhrat^{109/48} \rprtrat^{43/16}\!\!\mstarrat^{-9/8}
\rstarrat^{-27/16}\left(\frac{t}{\tfall}\right)^{15/8}.
\end{eqnarray}
If $\tlte < t_{\rm dyn}$, the assumption of blackbody emission
in equations (\ref{Loutflow}) and (\ref{tphnoedge}) is reasonable.
Equation (\ref{therm}) shows that this is generally true at early times
for $\mbh \lesssim 10^6\msun$ and any stellar pericenter distance, and
for $\mbh \lesssim {\rm few} \times 10^6\msun$ and $\rp \sim 3\rs$.
However, for more massive BHs ($\mbh \sim 10^7 M_\odot$), the assumption of
thermal equilibrium is probably poor for $t \gtrsim t_{\rm fallback}$,
when most of the stellar debris returns to pericenter.  Moreover, for
a given BH mass and stellar pericenter distance, the post-shock plasma
is not in thermal equilibrium after a time
\begin{eqnarray} \label{nontherm}
t_{\rm non-therm} &\sim& 0.5 \, \mbhrat^{1/3}\rprat^{47/30}\left(\frac{\mstarrat}{\rstarrat}\right)^{3/5} \, {\rm day} \nonumber \\ & \sim & 
10 \, \mbhrat^{-32/45} \rprtrat^{47/30}\left(\frac{\mstarrat}{\rstarrat}\right)^{3/5} \, {\rm day}.
\end{eqnarray}
Equation (\ref{nontherm}) shows that the assumption of thermal
blackbody emission is likely reasonable for a few weeks for $\mbh \sim
10^6 M_\odot$ and $\rp \sim \rt$.  For smaller $\rp$, thermal
equilibrium breaks down earlier, but events with $\rp \sim \rt$ are
predicted to dominate the rates and are thus in practice probably the
most important (Fig. 12 of \citealt{sq09}).  At times $t \gtrsim
t_{\rm non-therm}$, the temperature at the shock, $T_{\rm shock}$, is
tens to hundreds of keV, scaling as $\rprat^{-4}$ for mildly
relativistic fallback speeds.  The radiation emitted at the electron
scattering photosphere will be cooler than this by a factor of $\sim
(\rphes/\rl)^{2/3}$ due to adiabatic expansion.  Compton upscattering
also likely gives the radiation a power-law spectrum.  We will discuss
the effects of this non-blackbody emission in
\S\S\ref{xrayspectsection} and \ref{discussion}, but defer a detailed
calculation of the non-LTE spectrum to future work.

It is important to note that if the gas and radiation are not able to
thermalize close to the shock (i.e., for $t \gtrsim t_{\rm
  non-therm}$), they are unlikely to thermalize further out in the
outflow instead: the equilbrium photon number density falls with
radius, as $\rho$, but the equilibrium free-free emissivity falls
faster, as $\rho^{11/16}$, so there is even less time to come into
thermal equilibrium at larger radii.

\vspace{-0.2cm}
\subsection{Spectroscopic Calculations}\label{spectcalc}

The spectrum of the outflow will be imprinted with spectral lines
produced by the outer layers of gas between the photosphere and the
edge of the outflow at $\redge$.  The photosphere is initially
determined by the outflow itself ($\rsource=\rphes$) but at later
times as the outflow subsides and becomes optically thin, the
photosphere is set by the accretion disk, with $\rsource = \rdisk$.
The gas outside $\rsource$ absorbs photons released deeper in, through
photoionization and bound-bound transitions, creating absorption and
emission features.  The outer gas is highly ionized by the central
source, and maintains photoionization equilibrium so long as the
recombination time is shorter than the expansion time: $t_{\rm rec}
\sim (n_e\alpha_{\rm rec})^{-1} < t$. Here $\alpha_{\rm rec}$ is the
recombination coefficient, $\sim 2\times10^{-13}\cm^3\s^{-1}$ for
hydrogen at $30,000\K$ and typically larger for heavier species. 

Early on, most absorption takes place at $r \sim \rphes$, where the gas
is always in photoionization equilibrium, since
\begin{eqnarray}\label{photoeq_rphes}
  \frac{t_{\rm rec}(\rphes)}{t} & \sim & [n_e(\rphes)\alpha_{\rm rec}]^{-1} \nonumber \\
   & \sim & 10^{-5} \frac{f_{\rm out}}{f_{v}}\mbhrat^{5/3}\rprat^{5/2}\frac{\mstarrat}{\rstarrat}\trattenday^{-8/3} \, .
\end{eqnarray}
At late times, most of the mass in the outflow resides in a shell at
$r\sim\redge$, which dominates the absorption.  There, the gas falls
out of photoionization equilibrium after a few months to a few years,
since
\begin{eqnarray}\label{photoeq_redge}
  \frac{t_{\rm rec}(\redge)}{t} & \sim & [n_e(\redge)\alpha_{\rm rec}]^{-1} \nonumber \\
  & \sim & 3\times10^{-3} \frac{f_{v}^{3}}{f_{\rm out}}\frac{\mbhrat^{5/2}\rprat^{3/2}}{\mstarrat\rstarrat^{3/2}}\trattenday \, ,
\end{eqnarray}
when $\Delta r_{\rm shell} \sim \vwind\tfall$; $t_{\rm rec}$ increases
even more later when $\Delta r_{\rm shell} \sim 0.1\vwind t$.

We determine the ionization and opacity structure of the gas outside
$\rsource$ by performing photoionization calculations with version 08.00
of the publicly available code Cloudy, last described by
\citet{ferland}.  To determine the spectrum, we then post-process
Cloudy's output to account for Doppler shifts by the outward motion of
the gas.  We will only show results below for times at which $t_{\rm
  rec} \lesssim t$ so that photoionization equilibrium is a reasonable
approximation.

For calculating line profiles, it is useful to divide the outflow into
two parts: at times $t\lesssim\tedd$, when the outflow is being
continuously driven, there is a radially extended outflow from
$\rsource$ to $\sim \redge$.  Because $\redge$ is generally much
larger\footnote{\label{tedge}In \citet{sq09}, we describe an early
  $\sim$day-long phase for small $\mbh$ and small $\rp$ during which
  $\rphes \sim \redge$; although there may be interesting
  spectroscopic features during this phase, we focus on later times
  when the physics of the escaping photons is more secure.} than
$\rsource$, the line-of-sight velocities of the gas span a wide range,
which causes absorption lines to be strongly velocity-broadened.  At
times $t\gtrsim{\rm few}\times\tfall$, there is also a narrow, denser
shell at $r\sim\redge$, which contains most of the mass (because most
of the mass is unbound at $\sim \tfall$).  In the shell, thermal
broadening may dominate over velocity broadening.  We first qualitatively
describe the evolution of the spectrum produced by these two parts of
the outflow, and then explain in more detail how we calculate the
absorption and emission line profiles.

\vspace{-0.2cm}
\subsubsection{Three phases of evolution}\label{3phases}

For $\mbh\lesssim {\rm few} \, 10^7 \, M_\odot$, the fallback
rate is super-Eddington and the outflow is optically thick to electron
scattering for a few weeks to months after disruption.  Photons
released from $\rsource=\rphes$ with the blackbody spectrum in
equation (\ref{Loutflow}) (perhaps with an additional X-ray power-law
tail; \S\ref{xray}) photoionize the outer layers of gas between
$\rsource=\rphes$ and $\redge$, which have a density profile given by
equation (\ref{densityprofile}).  This gas spans a wide range in
radii, and so produces a spectrum of broad absorption lines, whose
profiles are described in \S\ref{velocbroad}; when $t\gtrsim{\rm
  few}\times\tfall$, there are also narrower absorption lines, described
in \S\ref{thermbroad}.  The emission lines are calculated as described
in \S\ref{emissionlines}.

At later times, the fallback rate diminishes and the electron
scattering photosphere of the outflow moves inward.  The entire
outflow becomes optically thin to electron scattering $\sim$ weeks to
months after disruption, but it continues to be driven until $\mdotfb$
falls below $\mdotedd$, which can be somewhat later ($\tedd$;
eq. \ref{tEdd}).  Deep inside the outflow, accretion onto the BH
proceeds via a thin disk, which emits a multicolor blackbody spectrum
peaking close to $\sim10^5\K$ \citep[described in][]{sq09}.  In our
calculations for these times, our input spectrum to Cloudy is the
spectrum of the accretion disk (whose size is $\rsource=\rdisk$).  The
outflow continues to span a wide range in radii, $\rsource=\rdisk
\lesssim r \lesssim \redge$, and so we calculate absorption line
profiles as in \S\ref{velocbroad}, with additional absorption lines
from the shell (\S\ref{thermbroad}) when appropriate.  The emission
lines are again calculated as in \S\ref{emissionlines}.

For $t > \tedd$, $\mdotfb < \mdotedd$ and the shocked gas at
pericenter can cool efficiently; essentially all of the gas thus
accretes through the disk rather than being blown away.  The
previously expelled material continues to expand out as a thin shell,
with a density given by equation (\ref{shell_rho}) and a radial
thickness $\Delta r_{\rm shell}$.  The shell is irradiated by the
blackbody emission from the accretion disk.  In many cases, the
outflow is no longer in photoionization equilibrium (see
eq. \ref{photoeq_redge}), but if there are places where the outflow
velocity is low ($f_v \sim 0.1$), the shell can be dense enough to
remain in equilibrium for up to several years.  In such cases, the
absorption lines are all {\it narrow} (\S\ref{thermbroad}), while the
emission lines are broad (\S\ref{emissionlines}) and very faint.

We now describe how we calculate the line profiles for the extended
part of the outflow and the shell \citep[see also related calculations
in][for line profiles in Wolf-Rayet star winds]{castor70}.
\vspace{-0.35cm}
\subsubsection{Velocity-broadened absorption lines}\label{velocbroad}

Cloudy outputs a table of absorption lines: each entry contains the
line frequency $\nu_0$, the species (element and ionization stage)
producing the transition, and the total optical depth $\tau_{\rm
  stat}$ through the (stationary) layer of gas.  Cloudy also outputs
the density distribution $n_{\rm species}(r)$ for each species, and
the temperature profile $T(r)$.  The stationary optical depth is
\begin{eqnarray}
\tau_{\rm stat} & = & \int_{\rsource}^{\redge} \!\! n_{\rm species}(r) \sigma_0(r) \, dr \\
& = & (\sigma_0 v_{\rm th}) \int_{\rsource}^{\redge} \!\! \frac{n_{\rm species}(r)}
{v_{\rm th}(r)} \, dr \\
& \equiv & (\sigma_0v_{\rm th}) I_{\rm species} \, ,
\end{eqnarray}
where $\sigma_0(r)$ is the cross section of the transition at line
center and $v_{\rm th}(r)$ is the thermal velocity of the gas.
Because the lines provided by Cloudy are thermally broadened, the
quantity ($\sigma_0 v_{\rm th}$) is independent of radius.

For simplicity, we assume that the gas flows out radially with a
spatially and temporally constant velocity, $v = \vwind$ (eq.
\ref{vwind}), superposed by small thermal motions. Since the gas is
optically thin at most frequencies, its temperature regulates to $T
\sim 10^4 - 10^5\K$, leading to thermal velocities $v_{\rm th} \sim
10-30\km\s^{-1} \ll \vwind \sim 0.01c-0.1c$.  Because the
photoionizing source---the electron scattering photosphere or
accretion disk---is spatially extended, its radiation originates from
impact parameters $b$ ranging from 0 (center of the source) to
$\rsource$ (edge of the source).  At a given impact parameter $b$, our
line of sight passes through gas moving at projected line-of-sight
velocities $\vlos = \vwind\sqrt{1-(b/r)^2}$,
where $r$ ranges from $\rsource$ to $\redge$.

A transition of frequency $\nu_0$ can absorb photons of rest frequency
$\nu$ at places in the wind where $\vlos$ satisfies $\nu =
\nu_0(1+\vlos/c)$.  
These locations are centered at radii
\begin{equation}
r_{\rm abs} = \frac{b}{\sqrt{1-(\vlos/\vwind)^2}}
\end{equation}
with a small spread along the line of sight, $\Delta \ell$, due to
random thermal motion of the gas:
\begin{equation}
\Delta \ell = r_{\rm abs} \left(\frac{v_{\rm th}(r_{\rm abs})}
{\vwind}\right)\left(\frac{r_{\rm abs}}{b}\right)^2  \, .
\end{equation}
Thus, for a given rest frequency and impact parameter, the optical
depth to a given transition is\footnote{We use $\tau_{\rm stat}$ and
  $I_{\rm species}$ because Cloudy does not output line opacities as a
  function of radius.}
\begin{equation}
\tau_b  \sim  \left.(n_{\rm species} \sigma_0 \Delta \ell)\right|_{r_{\rm abs}} 
 \sim \frac{\tau_{\rm stat} r_{\rm abs}^3 n_{\rm species}(r_{\rm abs})}{b^2 \vwind I_{\rm species}} \, .
\end{equation}
Approximating the source as a spherical isotropic emitter, we find the
overall transmitted power:
\begin{equation}\label{Ltrans}
\nu L_\nu^{\rm trans} = \nu L_\nu^{\rm source} \times \int_0^{\pi/2} 
2e^{-\tau_b} \sin\theta\cos\theta \, d\theta \, ,
\end{equation}
where $\sin\theta \equiv b/\rsource$.  When multiple transitions
contribute absorption at frequency $\nu$, we replace $\tau_b$ above
with the sum of their corresponding optical depths.  We account for
continuous absorption processes as well; these are generally
important only for $h\nu \gtrsim 0.3\,{\rm keV}$ since the gas is so
highly ionized.

\subsubsection{Thermally-broadened absorption lines}\label{thermbroad}

At times $t \gtrsim {\rm few} \times \tfall$, the bulk of the
previously expelled gas forms a thin shell at $r \sim \redge$, as
described above in \S\ref{basicprops}.  While velocity broadening is
the dominant broadening mechanism for absorption at $\rsource \lesssim
r \lesssim \redge$, the line-of-sight velocity's variation with impact
parameter at $r \sim \redge$ may become less than $v_{\rm th}$ after
at most a few months if the wind velocity is close to constant in time.
For example, at times when $\rsource=\rphes$,
\begin{displaymath}
\frac{\Delta \vlos}{v_{\rm th}} 
 \sim  \left(\frac{\rphes}{\redge}\right)^2\left(\frac{\vwind}{v_{\rm th}}\right) \nonumber
\end{displaymath}
\begin{equation}\label{deltavlos}
\sim 0.03\,\frac{f_{\rm out}^{2}}{f_v^{3}}\mbhrat^{16/3}\rprat^{11/2}\frac{\mstarrat^2}{\rstarrat^{2}}\!\!
\left(\frac{v_{\rm th}}{30\km\s^{-1}}\right)^{-1}
\!\!\trattenday^{-16/3} . 
\end{equation}
(The variation in $\vlos$ along the line of sight through the narrow
shell is even smaller than this, by a factor of $\Delta r_{\rm
  shell}/\redge$.)  Consequently, the random thermal motion may dominate
the broadening of absorption lines produced in the shell; such lines are
therefore {\it narrow}, with a linewidth $\Delta \nu = \nu_0(v_{\rm
  th}/c)$, and are blueshifted to $\sim \nu_0(1+\vwind/c)$.

If the wind speed varies in time by more than $\sim 0.1\%$, velocity
broadening instead dominates over thermal broadening, leading to wider and
shallower absorption lines.  The wind speed variation $\Delta \vwind$
could be substantial, e.g., $\sim 0.1\vwind$, but is highly uncertain, so
we consider thermal broadening in the shell as a lower limit.

We run separate Cloudy calculations for the narrow shells, which give
the thermally-broadened optical depth $\tau_{\rm th}(\nu)$ as a
function of $\nu$, which we blueshift by $\vwind/c$.  At times
$t\lesssim\tedd$, we multiply the velocity-broadened spectrum
described above by $e^{-\tau_{\rm th}(\nu)}$; at later times, we
multiply the continuum produced at $\rdisk$ by this factor.

\subsubsection{Emission lines}\label{emissionlines}

The gas at $r \gtrsim \rsource$ can also produce emission via
radiative recombination and radiative decay of (collisionally or
radiatively) excited atoms/ions.  Because of the large velocities in
the outflow, the gas is effectively optically thin to all of these
photons, even at the energies of resonance lines.  As a result, we can
observe emission from gas having line-of-sight velocities $\vlos \sim
-\vwind$ up to $\vlos \sim +\vwind(1-\rsource^2/\redge^2)^{1/2}$; we
approximate this with a Gaussian line profile centered on the rest
energy of the line, and slightly truncated on the red side.  This
approximation is based on analogy to Monte Carlo calculations of
emission from expanding Lyman alpha blobs \citep{verhamme}.  We treat
emission lines produced in the narrow shell the same way, since line
photons emitted from essentially any part of the shell can reach us.

The large velocities in the outflow imply that the resulting emission
lines are extremely broad; in most of our calculations, the emission
is so spread out that the lines will be orders of magnitude fainter
than the transmitted spectrum.  The overall emission $\nu L_\nu^{\rm
  emis}$ then consists of faint continuum and very broad emission
lines centered on the rest frequencies of the transitions.  The
observable spectrum is the sum of this emission and the transmitted
light from equation (\ref{Ltrans}), i.e.,
\begin{equation}
\nu L_\nu^{\rm out} = \nu L_\nu^{\rm trans} + \nu L_\nu^{\rm emis} \, .
\end{equation}

Although the physical processes are similar in tidal disruption events
and broad absorption line quasars (BAL QSOs), we note that tidal
disruption events have substantially smaller emission line equivalent
widths; this comparison is discussed further in \S\ref{discussion}.

\vspace{-0.5cm}
\section{Predicted Spectra}\label{results}

We now use the methodology of \S \ref{outflowprops} to calculate
spectra as a function of time due to the disruption of a solar-type
star, varying the BH mass and pericenter distance of the stellar
orbit. We assume solar abundances. We focus on solar type stars
because they are among the most abundant stars at $\sim 1-10$ pc in
galactic bulges, which is where most of the disrupted stars originate.
The three fiducial models we consider are: $\mbh=10^6\msun$ and
$\rp=3\rs$; $\mbh=10^6\msun$ and $\rp=\rt$; and $\mbh=10^7\msun$ and
$\rp=\rt$.  Our fiducial model for the outflow takes $f_v=1$ and
$f_{\rm out} = 0.1$, but later we vary these values.

To start, we assume that the outflow produces a thermal blackbody
spectrum that photoionizes the surrounding gas; in \S
\ref{xrayspectsection} we consider the effects of (harder) non-thermal
emission on the predicted spectra.  The thermalization time estimate
in \S \ref{sec:te} implies that thermal equilibrium in the outflow is
maintained for $t \lesssim 0.5$ and 10 days for $\mbh =10^6\msun$,
$\rp=3\rs$ and $\rp=\rt$, while it fails at later times; the outflow is
never in thermal equilibrium for $\mbh=10^7\msun$.  Given, however,
the uncertainties in the precise thermalization time we show thermal
outflow models for a range of timescales.

\begin{figure*}
\centerline{\epsfig{file=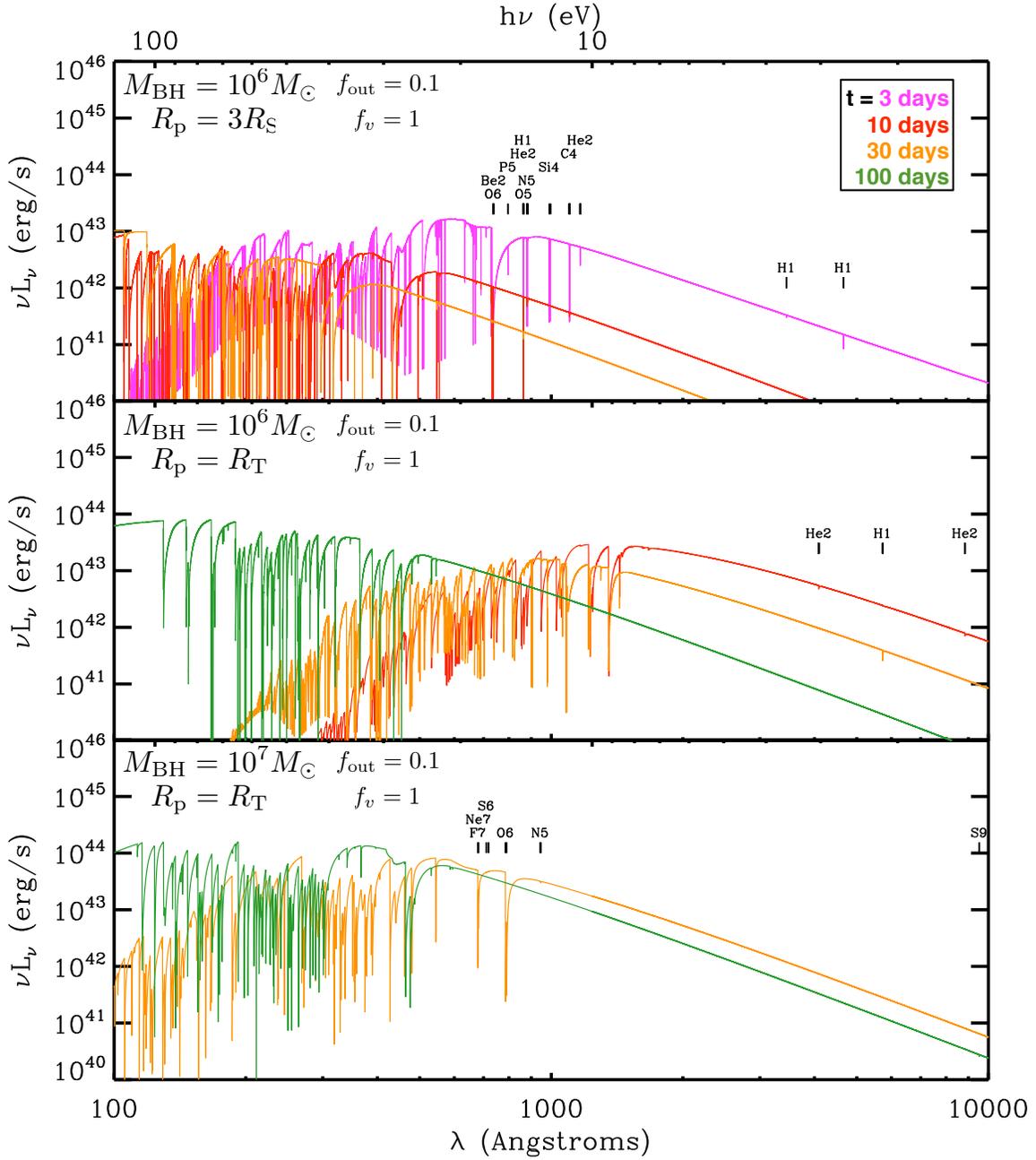, width=16cm}}
\vspace{-0.4cm}
\caption{Predicted spectra for our three fiducial tidal disruption
  flares at several different times after disruption.  Labeled tickmarks identify the blueshifted positions of the strongest long-wavelength lines.  A zoomed in view of the FUV region of the middle panel is in Figure \ref{zoomin_fout}.  At early times, the continuum emission in these calculations is produced by a super-Eddington outflow while at later times it is produced by the accretion disk close to the black hole.  These calculations assume that the outflow is able to thermalize completely (\S\ref{sec:te}); Fig. \ref{xray} shows results for incomplete thermalization.
  \label{fiducspect}}
\end{figure*}

\begin{figure*}
\centerline{\epsfig{file=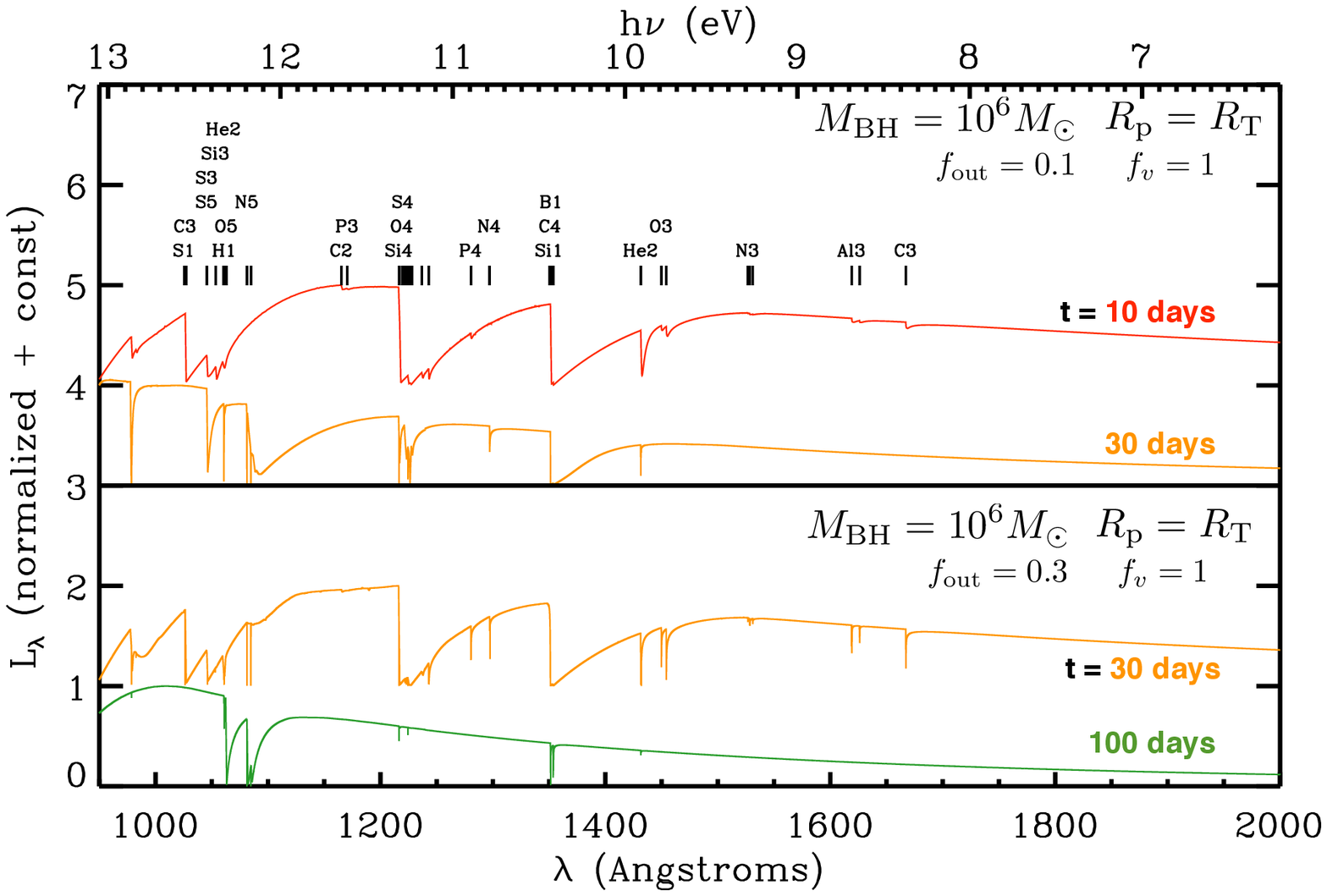, width=16cm}}
\vspace{-0.4cm}
\caption{Predicted spectra varying the mass-loss rate in the outflow
  (via $f_{\rm out}$; eq. \ref{mdotout}), focusing on the wavelength
  region $1000-2000\ang$.  Each spectrum is portrayed as $L_\lambda$,
  normalized by its maximum value on that wavelength range, with an
  added integer constant to offset curves for clarity.  Labeled tickmarks identify the   blueshifted positions of the lines.  Top panel: $\mbh=10^6\msun$,
  $\rp=\rt$, $f_{\rm out}=0.1$, $f_v=1$ at $t$ = 10 days and 30 days;
  bottom panel: same as top panel, except $f_{\rm out}=0.3$, and
  $t$ = 30 days and 100 days.  Lines tend to be stronger for larger mass outflow rates
  (larger $f_{\rm out}$), because the continuum emission has a lower temperature ($\tphes$).
  \label{zoomin_fout}}
\end{figure*}

Figure \ref{fiducspect} shows our predicted spectra at various times
after disruption, from the optical to the extreme ultraviolet
(EUV). Significant lines in the far ultraviolet (FUV) to optical are
labeled in the top and bottom panels.  The top panel of
Figure~\ref{zoomin_fout} shows a zoomed-in version of the $1000-2000
\ang$ region of the spectrum for our fiducial $\mbh=10^6\msun$ and
$\rp=\rt$ model.  For the latter, we plot the luminosity density
$L_\lambda$ normalized to its maximum value over the wavelength
interval $1000-2000\ang$ and vertically offset different curves for
clarity.

In Figures \ref{fiducspect} and \ref{zoomin_fout}, we show results
from $t \sim \tfall$, the peak of fallback and outflow, to when the
outflow shell falls out of photoionization equilibrium.\footnote{For
  $\mbh=10^6\msun$, $\rp=3\rs$ (top panel), the earliest time we
  depict is $t = 3$ days because $\rphes \sim \redge$ until $\sim 1$
  day (see footnote \ref{tedge}).}  (The outflow's photosphere
$\rphes$---where most absorption takes place while the fallback rate
is super-Eddington---is always in photoionization equilibrium; see
beginning of \S\ref{spectcalc}.)
Photoionization equilibrium in the shell typically fails when, or
somewhat before, the fallback rate reaches the Eddington rate. More
precisely, it fails at $\sim$ 30 days for $\mbh=10^6\msun$,
$\rp=3\rs$, $f_{\rm out}=0.1$; 100 days for $\mbh=10^6\msun$,
$\rp=\rt$, $f_{\rm out}=0.1$, and 200 days for $\mbh=10^6\msun$,
$\rp=\rt$, $f_{\rm out}=0.3$; while $\tedd \sim 30$, 400 and 400 days
for these three models respectively.  For $\mbh=10^7\msun$, $\rp=\rt$,
the shell of gas at $\sim \redge$ is never in photoionization
equilibrium, and so we do not include any contribution from the shell.

The spectral features visible in Figure \ref{fiducspect} are
exclusively absorption lines; the emission lines are so broadened by
the large range in line-of-sight velocity that they become
undetectable above the blackbody continuum.  The outflow as a whole
and the dense thin shell at $\redge$ typically both contribute to the
absorption lines.  For $\mbh=10^6\msun$, $\rp=3\rs$, the shell
dominates the absorption lines at $\lambda \gtrsim 800\ang$; the shell
and outflow as a whole both contribute at a wide range of wavelengths
for $10^6\msun$, $\rp=\rt$; and the shell does not contribute for
$10^7\msun$, $\rp=\rt$ because it is never in photoionization
equilibrium.  The absorption lines from the extended part of the
outflow are strongly blueshifted and typically (while $t < \tedd$)
very broad\footnote{The sawtooth shape of the absorption lines at
  early times in Figure \ref{fiducspect} is a result of our assumption
  of spatially constant velocity (see \S\ref{velocbroad}); a more
  realistic velocity gradient would shift some absorption from $\sim
  h\nu_0(1+\vwind/c)$ towards $h\nu_0$, leading to a less abrupt
  change in the spectrum at $\sim h\nu_0(1+\vwind/c)$.}, with
linewidths of $\sim \vwind \sim 0.01-0.1 \, c$.  Superimposed on these
broad lines are the thermally-broadened narrow lines produced in the
outer shell of gas, with linewidths of $\sim 10-30\km\s^{-1}$.

Most of the absorption lines are in the UV ($\lambda \lesssim
2000\ang$; $h\nu \gtrsim 10\,{\rm eV}$), with few features in the
optical.  The reason for this has two parts.  First, consider the
ionization parameter $U_{h\nu}$ for a species whose ionization energy
is $\xi_{\rm ion}=h\nu$: because the gas density is relatively low
while the incident spectrum 
is luminous and peaks at energy $h\nu_{\rm peak}\sim 10\eV$ or higher,
$U_{10\eV} \gg 1$
and so species having $\xi_{\rm ion} \lesssim 10\ev$ are almost fully
ionized.  (For example, the ionization parameter for hydrogen is
typically $U_{\rm H} \sim 10^3-10^5$, and hydrogen's neutral fraction
is typically $10^{-8}-10^{-10}$.)  Secondly, most atoms/ions are in
the ground state: the radiative decay rates are fast compared to the
rates of photoionization and collisional excitation.
As a result, almost all of the species
present in the flow have $\xi_{\rm ion} \gtrsim h\nu_{\rm peak}
\gtrsim 10\ev$, and members of those species are in the ground state.
Transitions from the ground state have energies similar to $\xi_{\rm
  ion}$, which is $\gtrsim 10\ev$, so most spectral lines have
energies $\gtrsim 10\ev$, i.e., in the UV rather than optical.  In
particular, some common prominent lines in the UV at $\lambda >
1000\ang$ are He II ($1640\ang$), C IV ($1548 + 1551 {\rm \AA}$), Si
IV ($1394 + 1403 {\rm \AA}$), O IV ($1400 + 1401 + 1405 + 1407 {\rm
  \AA}$), N V ($1239 + 1243 {\rm \AA}$), Lyman $\alpha$ ($1216 {\rm
  \AA}$), and O VI ($1032 + 1038 {\rm \AA}$).  These are similar to
the lines observed in the spectra of BAL QSOs---this is not surprising
given that the physical conditions are similar.  When the density is
high enough for optical lines to be present
(because $U$ and $h\nu_{\rm peak}$ are lower), 
these optical lines are mostly the lines of H I and He II.

Figure \ref{fiducspect} shows that the minimum energy/maximum wavelength of
the absorption lines in the optical--UV depends both on time and on the
parameters of the tidal disruption (e.g., $\mbh$ and $\rp$).  For
example, the typical absorption lines shift to shorter wavelengths at
later times.  The physical origin of these dependencies can be
understood as follows.  For all times and outflow parameters that we
consider, $U_{h\nu}$ in the outflow is $\gg 1$ for $h\nu \lesssim
h\nu_{\rm peak}$, and so the approximate minimum
energy\footnote{High-$n$ lines of hydrogen at wavelengths of tens of
  microns are also typically optically thick due to $l$-mixing
  collisions with ions.} of significant absorption lines present in
the spectra is $\sim h\nu_{\rm peak}/{\rm few}$.  So long as the
outflow is optically thick (most times depicted), the peak energy of
the incident spectrum is set by the temperature of the electron
scattering photosphere $\tphes$.  We can thus understand the variation
in the typical energy of spectral lines by considering the scalings
for $\tphes$ in equation (\ref{tphnoedge}).  For example, at a fixed
time after disruption, a model with $\mbh=10^6\msun$, $\rp=3\rs$ has a
hotter electron scattering photosphere than a model with
$\mbh=10^6\msun$, $\rp=\rt$. This is why the spectra for the former
situation have fewer low-energy lines in Figure \ref{fiducspect};
$\mbh=10^7\msun$, $\rp=\rt$ is yet hotter at fixed time, and so has
even fewer longer wavelength lines.

These arguments also help to explain the time evolution of the spectra
in Figure \ref{fiducspect}.  During the optically thick phase
(\S\ref{3phases}), the continuum radiation produced by the outflow
becomes harder with time as $\rphes$ moves inward, while the
luminosity remains high.  As a result, the lines present in the
spectrum tend to have shorter wavelengths (higher energies) at later
times.  Once the outflow becomes optically thin, the continuum
spectrum is $\sim 30\eV - 100\eV$ emission from the accretion disk,
and most absorption lines have $h\nu \gtrsim 30\eV$ and remain very
broad (\S\ref{velocbroad}).  In some cases, narrow lines produced in
the shell (\S\ref{thermbroad}) are superimposed on these broad lines.
The lines would become purely narrow after $\tedd$ because nearly all
of the outflow is then in a thin shell, but for our fiducial outflow
parameters, the outflow falls out of photoionization equilibrium
before $\tedd$.

\begin{figure*}
  \centerline{\epsfig{file=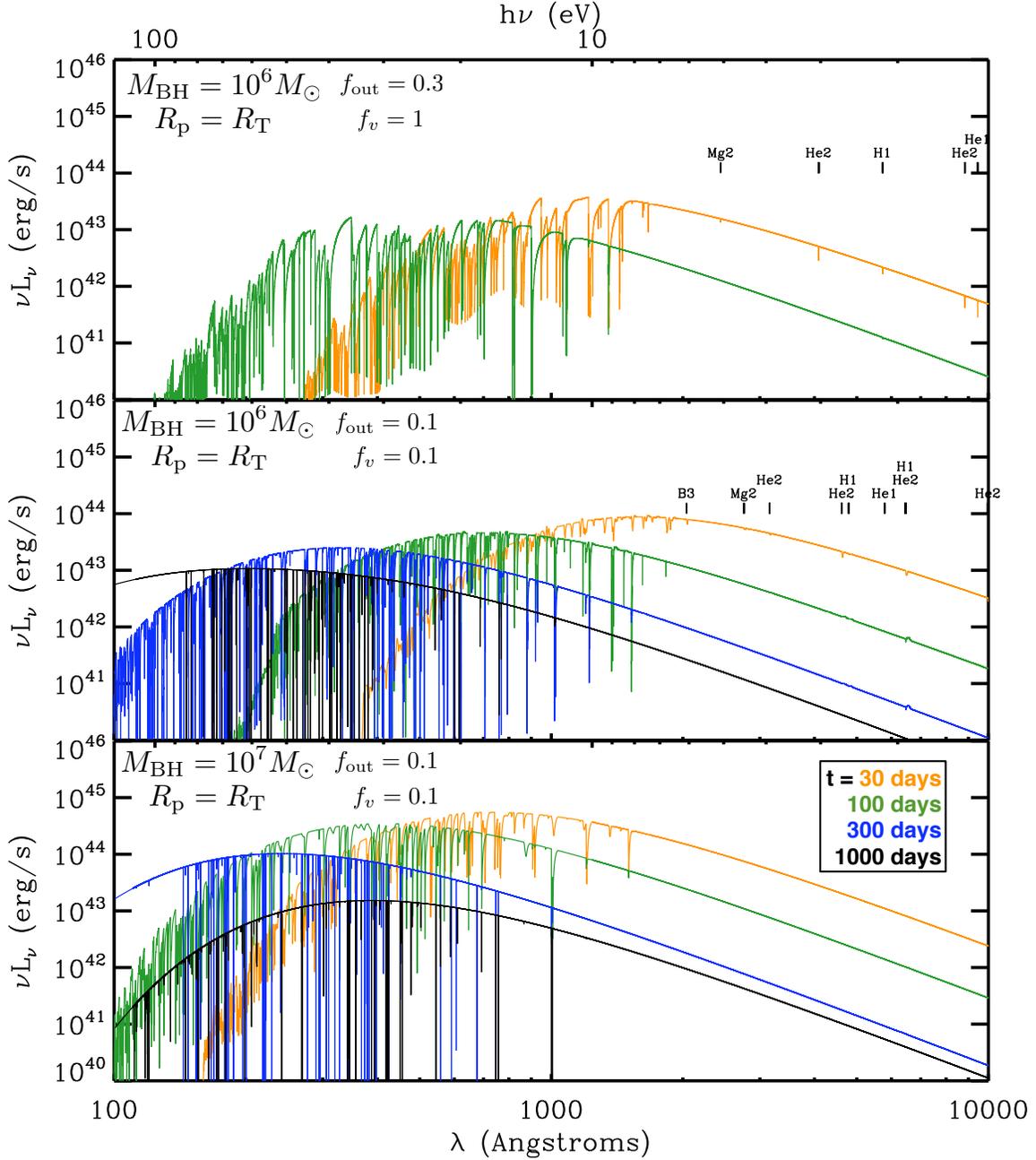, width=16cm}}
  \vspace{-0.4cm}
\caption{Predicted spectra showing the effects of varying the mass-loss rate (via $f_{\rm out}$; eq \ref{mdotout}) and outflow speed (via $f_v$; eq. \ref{vwind}) in the super-Eddington wind.  Labeled tickmarks identify the blueshifted positions of the strongest long-wavelength lines.   Slower and/or higher mass-loss rate winds have softer continuum emission and are thus more likely to produce optical or near-UV lines (compare these results with Fig. \ref{fiducspect}).  Top panel:  $\mbh=10^6\msun$, $\rp=\rt$, $f_{\rm out} = 0.3$, $f_{v}=1$, $t=30$ days, 100 days; middle panel: $\mbh=10^6\msun$, $\rp=\rt$, $f_{\rm out} = 0.1$, $f_{v}=0.1$, $t=30$ days, 100 days, 300 days, 1000 days; bottom panel: $\mbh=10^7\msun$, $\rp=\rt$, $f_{\rm out} = 0.1$, $f_{v}=0.1$, $t=30$ days, 100 days, 300 days, 1000 days.   A zoomed in view of the FUV region of the middle panel is in Figure \ref{zoomin_fout}.
  \label{fvfout}}
\end{figure*}

As mentioned in \S\ref{basicprops}, numerical simulations suggest that
the velocity and density of the outflow will vary with latitude, with
higher speed outflows along the pole relative to the equator (e.g.,
\citealt{ohsuga05}).  To consider how the spectrum of a tidal disruption
event may vary with viewing angle, Figure \ref{fvfout} shows spectra
for different values of $f_{\rm out}$ and $f_v$ (eqs. [\ref{mdotout}]
and [\ref{vwind}]).  Significant NUV/optical lines are labeled; the
lower panel of Figure \ref{zoomin_fout} highlights the
$\lambda=1000-2000\ang$ part of the spectrum for the $f_{\rm out} =
0.3$, $f_v = 1$ model.
Because the density at $r\sim\redge$ is larger if the outflow is
slower (smaller $f_v$) the shell remains in photoionization
equilibrium longer for the models with $f_v=0.1$ in Figure
\ref{fvfout}, for 4000 days ($\mbh=10^6\msun$, $\rp=\rt$) and 1000
days ($\mbh=10^7\msun$, $\rp=\rt$).

The ionization parameter $U_{h\nu}$ is large for all of these
variations about our fiducial models.  It is thus again the
temperature of the continuum radiation $\tphes$ that determines the
approximate minimum energy of the spectral lines.  For $f_{\rm
  out}=0.3$, $\tphes$ is lower than for $f_{\rm out}=0.1$, and so
there are more and deeper FUV-optical lines when $f_{\rm out}$ is
larger (compare the top panel of Figure \ref{fvfout} with the middle
panel of Figure \ref{fiducspect}, or the two panels of Figure
\ref{zoomin_fout}); for $f_{\rm out}=0.01$ (not shown), there are
virtually no lines with $\lambda \gtrsim 1500\ang$.  Similarly,
$\tphes$ is lower for lower outflow velocities.  For $\mbh=10^6\msun$,
$\rp=\rt$, $f_v=0.1$, there are in fact many optical absorption lines
(mostly hydrogen Balmer and He II) early on; for $\mbh=10^7\msun$,
$\rp=\rt$, $f_v=0.1$, there are many lines at $\sim1000-2000\ang$,
though no lines in the optical.  Additionally, H$\alpha$ and He II
$\lambda$6560 (blended together) can be seen in emission at late times
for $\mbh=10^6\msun$, $\rp=\rt$, $f_v=0.1$.  Since the outflow
velocity is lower, the density at $r\sim\redge$ is larger and so
recombination happens more frequently, leading to emission lines;
furthermore, the lower outflow velocity produces less broadening and
so the emission lines are brighter above the blackbody continuum.

\subsection{Implications of an X-ray Power-law}\label{xrayspectsection}

\begin{figure*}
\centerline{\epsfig{file=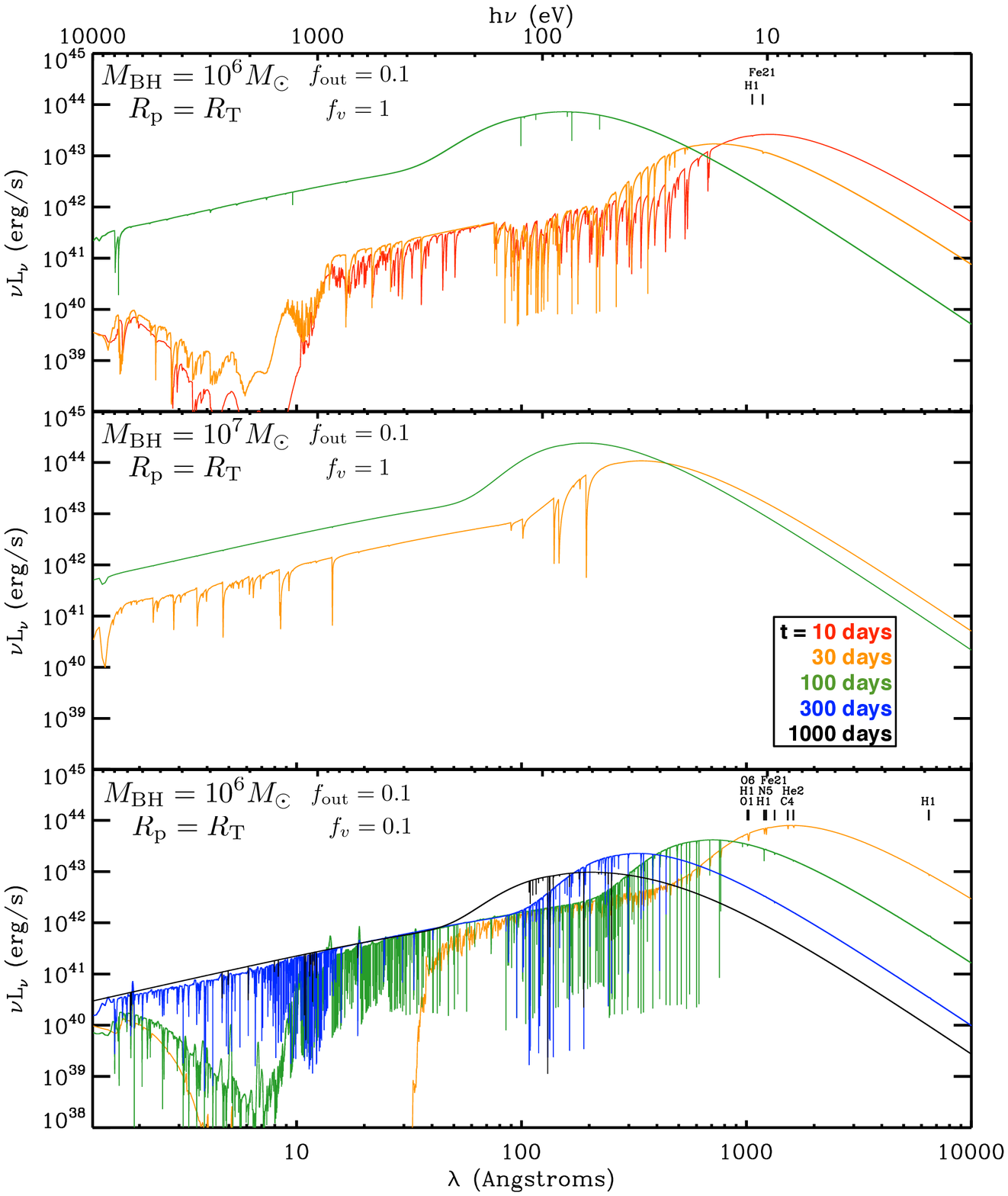, width=16cm}}
\vspace{-0.4cm}
\caption{Predicted spectra, including the presence of an X-ray power-law tail with photon index $\Gamma=3$ that carries 10\% of the blackbody luminosity (appropriate when the shock at pericenter does not have time to thermalize completely).  Optical and UV lines are mostly absent because the gas is so highly ionized (cf. pure blackbody continuum in Figs. \ref{fiducspect} and \ref{fvfout}), although there is typically significant absorption in the EUV and soft X-rays.  When $\Gamma=2$, there are even fewer lines and little continuum absorption.
\label{xray}}
\end{figure*}

The candidate tidal disruption events discovered in the ROSAT All-Sky
Survey and GALEX Deep Imaging Survey all show soft X-ray spectra
\citep{komossa, gezari08, gezari09}.  However, by analogy to the observed
spectra of AGN, it is possible that some tidal disruption spectra will
contain a high-energy power-law tail extending from the peak in the
blackbody continuum to hard X-rays.  Such an X-ray power-law component
might also be produced
as a result of incomplete thermalization at the shock at pericenter
where matter falls back to the BH (\S\ref{sec:te}).  To consider the
observational effects of such X-rays, we carried out Cloudy
calculations using an input spectrum that consists of the blackbody
spectrum described in \S\ref{3phases} plus a power-law tail that has
10\% of the blackbody luminosity, a photon index of 3 ($\nu L_\nu
\propto \nu^{-1}$), and that begins at the frequency where its
emission equals the blackbody emission.  (Such a spectrum may be
appropriate for partial thermalization; in a moment, we consider the
even harder spectrum expected from Compton equilibrium if the shock is
not thermalized.)

Figure \ref{xray} shows the 10 keV to 1$\mum$ spectra for several of
our models including this X-ray power-law: $\mbh=10^6\msun$,
$\rp=\rt$, $f_v=1$ (top panel); $\mbh=10^7\msun$, $\rp=\rt$, $f_v=1$
(middle panel); and $\mbh=10^6$, $\rp=\rt$, $f_v=0.1$ (bottom panel).
Due to the different peak energies of the blackbody components, the
actual power-law luminosity above 1 keV varies for the different
models as a function of time, with $\nu L_\nu(>1\kev)/L_{\rm bol} =
10^{-4}-2\times 10^{-2}$ (top panel), $6\times 10^{-3}-10^{-2}$
(middle panel), and $5\times 10^{-5}-10^{-2}$ (bottom panel).  These
models are thus reasonably conservative in terms of the contribution
of the X-ray emission to the bolometric luminosity.  Nonetheless, the
presence of the hard X-ray emission significantly changes the
resulting spectra.

The hard incident spectrum photoionizes the gas to a higher degree
than the pure blackbody incident spectrum.  Species that have
ionization energies $\xi_{\rm ion}<10\eV$ are thus more scarce, and so
the optical absorption lines seen in the previous section disappear
(except for $\mbh=10^6\msun$, $\rp=\rt$, $f_v=0.1$, where H$\alpha$
still has optical depth $\sim1$).  Most of the FUV lines disappear as
well; the Ly$\alpha$ line and a few others can still be faintly
visible early on for $\mbh=10^6\msun$, especially if $f_v \sim 0.1$.
Fe XXI $\lambda$1354 is sometimes the longest wavelength UV/optical
line.

So long as the outflow is optically thick to electron scattering, the
X-ray tail can show many absorption features.  These include both
continuum absorption and individual absorption lines.  Figure
\ref{xray} shows that for $\mbh=10^6\msun$, $\rp=\rt$, there is a deep
continuum absorption trough, extending from $\sim$ 1 keV up to $\sim
5$ keV;
for $f_v=0.1$ the outflow is denser and the trough extends down to
$\sim 0.3$ keV.  This feature is somewhat weaker for $\mbh=10^6\msun$,
$\rp=3\rs$ and absent for $\mbh=10^7\msun$.  There are also many
absorption lines superposed on the power-law tail and absorption
trough, provided by highly ionized Ar, Ca, Fe, Mg, Mn, Ni, and Si
(among others). The specific lines and line strengths vary
significantly between the different models, so it is difficult to
predict exactly which lines will be in the spectrum.  As the electron
scattering photosphere moves inward and becomes hotter, Figure
\ref{xray} shows that the X-ray luminosity actually increases in time
given our assumption of a fixed power-law with a photon index of 3.
Once the outflow becomes optically thin to electron scattering and the
accretion disk provides the incident spectrum with $h\nu_{\rm peak}
\sim 0.1\kev$ (typically after a few months), the gas is so highly
ionized that most of the absorption and emission features in the
spectrum disappear.  

If instead of $\nu L_\nu \propto \nu^{-1}$, the X-ray power-law tail
has a flatter spectrum with a photon index of 2 (extending up to $\sim 100$
keV), the gas is even more highly ionized. There are thus fewer X-ray
lines (though still a significant number) and little continuum
absorption.  There are no lines at all in the FUV/optical (apart from
very faint Ly$\alpha$ and Fe XXI $\lambda$1354 for $f_v=0.1$).

\vspace{-0.5cm}
\section{Supernova Rates in Galactic Nuclei}\label{SNrates}

\begin{figure}
\centerline{\epsfig{file=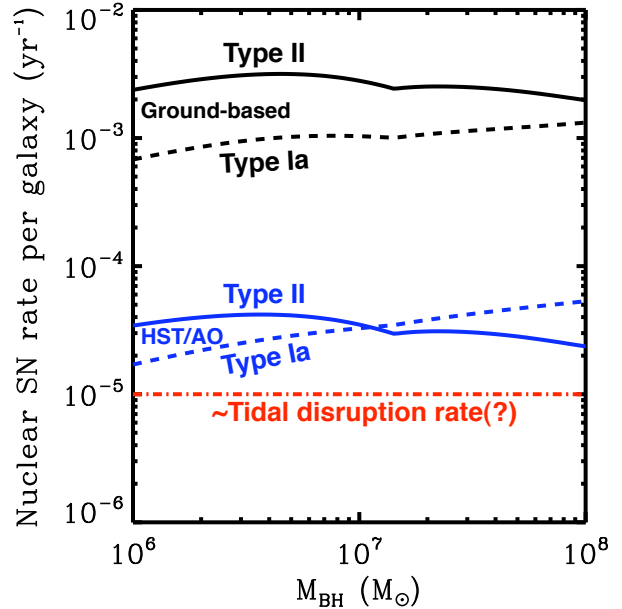, width=9cm}}
\vspace{-0.4cm}
\caption{Supernova rates within 0.5'' (black) and 0.05'' (blue) of the galactic nucleus at
  $z=0.1$, for Type II supernovae (solid) and Type Ia supernovae
  (dashed); 0.5'' is the typical astrometric
  accuracy of ground-based transient surveys  (J. Bloom, personal communication) while 0.05'' is the
  spatial resolution of HST or ground-based adaptive optics.
  If the rate of tidal disruption per galaxy is
  $10^{-5}\yr^{-1}$ \citep[red dot-dashed line]{donley},
  then nuclear supernovae 
  will outnumber tidal disruption events by about two orders
  of magnitude for ground-based optical transient surveys.  For HST/AO, the nuclear supernova rate will be comparable to the tidal disruption rate, reducing the contamination dramatically.
  \label{SNratesfig}}
\end{figure}

Tidal disruption flares are expected to be similar to supernovae
  in their overall brightness and timescale.  Supernovae thus
represent a significant source of contamination when trying to
discover and study tidal disruption flares in optical transient
surveys.  In particular, tidal disruption flares may be confused with
supernovae that appear coincident with the galactic nucleus within the
spatial resolution of the observations.  To quantify this source of
confusion, we estimate the rate of Type II and Type Ia supernovae that
take place within a distance $R_{\rm res}$ of a galaxy's nucleus.  To
do so, we first estimate the disk and bulge stellar mass within
$R_{\rm res}$ as a function of BH mass.

For a BH of mass $\mbh$, the stellar mass of the bulge is $M_{\ast,B}
\sim 700\mbh$ \citep{haring04}.  We estimate the stellar mass of the
disk using the ratio of bulge ($B$) to total mass ($T$) from Figure 3
of \citet{hopkins09}, which is based on data from \citet{balcells}: if
$M_{\ast,B}<10^{10}\msun$, $B/T \sim 0.1$; for larger
masses, $B/T \sim 3 \log (M_{\ast,B}/10^{10}\msun)-2.9$
(where total mass $M_{\ast,T}$ = disk mass $M_{\ast,D}$ + bulge mass $M_{\ast,B}$).  We determine
the half-light radius of the disk $R_{1/2,D}$ using the results from
SDSS in \citet{shen03} (their eq. 18) and convert to scale radius
via $R_{{\rm scale},D} = R_{1/2,D}/1.68$.  We then find the mass of the
stellar disk inside $R_{\rm res}$ by assuming that the surface density
profile is exponential:
\begin{equation}
M_{\ast,D}(< R_{\rm res}) \simeq M_{\ast, D}\int_0^{R_{\rm res}/R_{{\rm scale},D}}xe^{-x}\,dx \, .
\end{equation}
Similarly, we calculate the half-light radius of the bulge $R_{1/2,B}$
hosting a given BH using\footnote{Note the erratum in \citet{shen03}'s
  Table 1, so that $b=2.88\times 10^{-6}$.} equation (17) in
\citet{shen03} and convert it to the scale
radius of the de Vaucouleurs profile using $R_{{\rm scale},B} =
R_{1/2,B}/3461$ \citep{binneymerrifield}.  We then find the mass of
the bulge inside $R_{\rm res}$ using
\begin{equation}
M_{\ast,B}(< R_{\rm res}) \simeq \frac{M_{\ast,B}}{20160}\int_0^{R_{\rm res}/R_{{\rm scale},B}}xe^{-x^{1/4}}\,dx \, .
\end{equation}

The rate of Type II supernovae depends on the instantaneous star
formation rate (SFR) in the galaxy.  For galaxies that are actively
forming stars, the galaxy-integrated SFR as a function of galaxy
stellar mass ($M_\ast=M_{\ast,D}+M_{\ast,B}$) is $\log SFR = 0.67 \log
(M_\ast/10^{10} M_\odot) - 6.19$, where $SFR$ is in $\msun\yr^{-1}$
\citep{noeske07}\footnote{We extend the relation down to $M_\ast\sim
  10^9\msun$, and we suppose that the relation is proportional to
  $(1+z)^{3.4}$, using the redshift dependence of the
  volume-integrated star formation rate in \citet{yuksel08} and in
  agreement with the approximate redshift dependence reported by
  \citet{noeske07}.}.  Assuming that the spatial distribution of star
formation tracks stellar mass, the SFR within $R_{\rm res}$ is
$SFR(<R_{\rm res}) \sim [M_{\ast,D}(<R_{\rm res})/M_{\ast,D}] \times
SFR$.  The nuclear Type II supernova rate is then \begin{equation}
  \Gamma_{\rm II}(<R_{\rm res}) \simeq 10^{-2}f_D \frac{SFR(<R_{\rm
      res})}{\msun\yr^{-1}} \yr^{-1}. \label{type2}
\end{equation}   To account for the fact that only disk galaxies typically form stars, equation (\ref{type2}) includes a multiplicative factor given by the fraction of disk galaxies at the given $M_\ast$,
\begin{equation}
 f_D \simeq 1- \frac{dn_E/d \ln M_\ast}{dn_{\rm tot}/d \ln M_\ast} \, 
\end{equation}
where $dn/d \ln M_\ast$ is the number density of elliptical galaxies
($E$) or all galaxies (tot) from \citet{bernardi10}
locally\footnote{Parameters for mass functions of ellipticals and
  total are from Table B2 in \citet{bernardi10}, defining ellipticals
  by a concentration index larger than 2.86 as recommended by these
  authors.} and \citet[][their Table 3]{drory09} at higher redshift.
With this factor of $f_D$, our estimate of the supernova rate per
galaxy statistically takes into account that some systems of a given
$M_\ast$ are already passive and thus will not have significant
numbers of Type II supernovae.  

Following recent galaxy integrated results, we estimate the rate of
nuclear Type Ia supernovae given both the nuclear stellar mass and the
nuclear star formation rate:
\begin{eqnarray}
\Gamma_{\rm Ia}(<R_{\rm res}) &\simeq& A \left(\frac{[M_{\ast,B}+M_{\ast,D}](<R_{\rm res})}{10^{10}\msun}\right) \\ \nonumber
& + & B \left(f_D\frac{SFR(<R_{\rm res})}{10\msun\yr^{-1}}\right) \, ,
\end{eqnarray}
where $A = 4.4\times10^{-4}\yr^{-1}$ and $B=2.6\times10^{-2}\yr^{-1}$ \citep{scannapieco05}.

The typical accuracy to which ground-based optical transient surveys
can determine the location of a transient is $\sim 0.5$'' (J. Bloom,
personal communication), which corresponds to $R_{\rm res} = 0.9$ kpc
at $z=0.1$ and $R_{\rm res} = 4.1$ kpc at $z=1$.  Figure
\ref{SNratesfig} shows the Type II and Type Ia supernova rates per
galaxy within $0.9$ kpc of the galactic nucleus, as a function of BH
mass; this choice of $R_{\rm res}$ corresponds to a ground-based
survey observing at $z \simeq 0.1$.  Predicted tidal disruption rates
depend on BH mass and galaxy structure and typically range from $\sim
10^{-6}-10^{-3}$ yr$^{-1}$ per galaxy.  Using candidate detections in
the ROSAT All-Sky Survey, \citet{donley} estimated a rate of $\sim
10^{-5}$ tidal disruptions per year per galaxy.  For the latter,
Figure \ref{SNratesfig} shows that nuclear supernova rates are
typically several orders of magnitude larger than tidal disruption
rates.  High resolution photometry with the Hubble Space Telescope
(HST) or ground-based adaptive optics (AO) can decrease the nuclear SN
rates to ${\rm few}\times 10^{-5} \, {\rm yr^{-1}}$ per galaxy, reducing
confusion to order unity.  We discuss the implications of these
estimates of nuclear supernova rates in the next section.

\vspace{-0.6cm}
\section{Discussion}\label{discussion}

We have calculated the spectroscopic signatures of outflows produced
by super-Eddington accretion during the tidal disruption of stars by
massive black holes.  Although there are some uncertainties in the
continuum emission, mass outflow rate and kinematics, we find a number
of reasonably robust conclusions: 1) the spectrum will show strong
absorption lines that are blueshifted relative to the host galaxy, 2)
the absorption lines will typically be very broad ($\sim 0.01-0.1 \,c$),
though often with a narrow thermally-broadened ($\sim 30\km\s^{-1}$)
component that dominates as the outflow subsides and becomes a thin
shell at later times,
3) if the continuum spectrum is largely a blackbody, the absorption
lines will be most prominent at UV wavelengths (e.g., C IV, Si IV, O
IV, N V, Lyman $\alpha$ and O VI).  In addition, if there is a lower
velocity component to the outflow, which is plausible based on
simulations of radiatively inefficient accretion
\citep[e.g.,][]{ohsuga05}, we find that there will also be optical
absorption (and possibly emission) lines, in particular H$\alpha$,
H$\beta$, He II $\lambda$6560, and He II $\lambda$4860
(Fig. \ref{fvfout}).
The optical/UV lines will, however, be largely absent from the
spectrum if the continuum emission is harder or contains an X-ray
power-law tail (either because of non-thermal processes or incomplete
thermalization)---the gas would then be too highly ionized.  In that
case, the dominant absorption lines are in the soft X-rays
(Fig. \ref{xray}).  This highlights the importance of X-ray
observations of tidal disruption events coeval with optical and UV
spectroscopic observations, in order to properly interpret the
presence or absence of optical/UV lines.

Having summarized our key results, we now describe several
uncertainties in our spectroscopic predictions.  Our calculations
assume a blackbody spectrum of photons released from the electron
scattering photosphere of the outflow, with an `optional'
phenomenologically motivated X-ray power-law tail.  We find, however,
that when gas returning to pericenter shocks at $\sim 2 \rp$, the
thermalization between the photons and gas is a strong function of BH
mass, stellar pericenter distance, and time since disruption (\S
\ref{sec:te}).  The thermalization is often incomplete.  This would
harden the continuum spectrum, eliminating many of the optical-UV
spectroscopic signatures of tidal disruption events.
Moreover, the prominent optical continuum emission predicted by
\citet{sq09} from tidal disruption outflows requires reasonable
thermalization in the post-shock plasma.  Such thermalization is the
most likely for tidal disruptions around lower mass BHs with $\mbh
\lesssim {\rm few} \, 10^6 M_\odot$ and at early times after
disruption, $t \lesssim 2$ weeks (eq. \ref{therm}).  This estimate
assumes spherical fallback and is thus somewhat conservative, since
stellar debris is focused into a thin stream as it falls back after
disruption \citep{kochanek94}; a higher density at the shock would
lead to a longer period of thermal blackbody emission, perhaps up to a
month for $\mbh \sim 10^6 M_\odot$.  Nonetheless, these results
emphasize the importance of high cadence observations with rapid
follow-up in optical searches for tidal disruption flares.
In the future, observing the spectrum of a tidal disruption event
change as the outflow falls out of thermal equilibrium would provide
strong constraints on the physics of radiation-dominated shocks, which
are important in other astrophysical environments such as shock
break-out in supernovae.

Another uncertainty in our spectroscopic predictions is related to the
mass loading and speed of the outflow (parameterized by $f_{\rm out}$
[eq. \ref{mdotout}] and $f_v$ [eq. \ref{vwind}] in our calculations),
and the outflow's geometry.  Radiation hydrodynamic simulations of
super-Eddington black hole feeding suggest reasonable values for
$f_{\rm out}$ and $f_v$: \citet{ohsuga05}'s simulation at
$100\mdotedd$ obtains $f_{\rm out}\sim 0.1$ and $f_v \sim 1$ (our
fiducial values). \citet{ohsuga07}'s similar simulations of feeding at
$\mdotedd - 300\mdotedd$ imply $f_{\rm out} \sim 0.1 - 0.4$ (depending
on viscous parameter $\alpha$) and outflow velocities of $\sim
0.1c-0.3c$; \citet{takeuchi09}'s similar simulations at $100\mdotedd$
and $300\mdotedd$ imply $f_{\rm out} \sim 0.8$.  \citet{dotan_shaviv}
calculate a super-Eddington accretion model in which instabilities
make the gas inhomogeneous, and find that $f_{\rm out} \sim 0.5 - 0.7$
for feeding rates $5\mdotedd - 20\mdotedd$.  We note that
\citet{lodato_rossi} calculate tidal disruption light curves using our
super-Eddington outflow model (eqs. [\ref{rphes}] - [\ref{tphnoedge}])
with \citet{dotan_shaviv}'s $f_{\rm out}$ results and a more detailed
model for $\mdotfb(t)$ \citep{lodato09}; they find similar results to
those in \citet{sq09} (optical luminosities are similar, and emission
from the outflow lasts a factor of few times longer).  In
\S\ref{results}, we used fiducial values of $f_{\rm out} \sim 0.1$ and
$f_v \sim 1$ and showed the results of varying those parameters:
larger mass outflow fractions lead to more absorption lines at longer
wavelengths, and slower outflow velocities lead to narrower absorption
lines, the presence of optical lines, and even the presence of some
emission lines.  In reality, the mass-loading and outflow kinematics
may vary with viewing angle; the geometry of the outflow is uncertain,
and we do not capture this effect with our simple spherical
models. \citet{Ohsuga_Mineshige07} and \citet{takeuchi09} find outflow
opening (half) angles of $\sim 30$ degrees, suggesting that from some
viewing angles, accretion disk and outflow may both be visible.

\vspace{-0.4cm}
\subsection{Observational Prospects}

The prominent UV lines predicted here are challenging to observe for
several reasons.  First, the UV emission can be obscured by dust,
particularly along lines of sight through the host galaxy's disk;
however, observations of similar absorption lines in BAL QSO spectra
suggest that at least some lines of sight will have low obscuration.
Second, observations at extreme UV wavelengths ($1000\ang \lesssim
\lambda \lesssim 100\ang$), where most of the predicted spectral
features lie, must take place from space and are technically
difficult.  Encouragingly, the Space Telescope Imaging Spectrograph
and the Cosmic Origins Spectrograph aboard the Hubble Space Telescope
(HST) should be able to observe spectroscopic features like those
predicted here in the far UV.

The optical to X-ray spectra predicted here apply to tidal disruption
flares having early-time super-Eddington outflows, which requires
$\mbh \lesssim {\rm few} \times 10^7 M_\odot$.  To detect such flares
in the first place using optical transient surveys, significant
contaminants such as variable active galactic nuclei (AGN) and
supernovae must be excluded.  Luminous AGN fueled by other means are
$\gtrsim 10^{3}$ times more common than tidal disruption events and
could in principle produce optically bright `flares.'
Typically, however, AGN show optical emission lines in their spectra
(like the Balmer lines and [O III]).  We have shown that during the
super-Eddington outflow phase, tidal disruption events are unlikely to
show such optical lines; furthermore, if tidal disruption events do
show optical lines, Balmer and He II lines are the only reasonable
candidates (perhaps also very faint He I and S IX), because the
densities are too high for collisionally excited lines like [O III]
which almost always appear in AGN spectra.  These conclusions about
Balmer versus [O III] lines also hold during the later phase of a
tidal disruption event (described in \citealt{sq09}) in which the
accretion disk irradiates the unbound stellar debris.

BAL QSOs (a subset of quasars) are more physically and
spectroscopically similar to tidal disruption events, both involving a
bright central continuum source driving an outflow, which gives rise
to blueshifted UV absorption lines.  However, BAL QSOs show strong
emission lines while tidal disruption events likely will not, for the
following two-part reason, which draws on \citet{murray95}'s
theoretical work on BAL QSOs.  First, the tidal disruption outflow is
typically much more ionized than a BAL QSO wind, so optical depths for
resonance lines in tidal disruption events are at most comparable to,
and are often much less than, optical depths in BAL QSOs.  (The
optical depth to true absorption---rather than scattering---by
resonance transitions is typically much less than one in both cases;
the emission is thus effectively thin and so the smaller optical depth
for tidal disruption outflows implies less emission.)  Secondly, BAL
QSO winds are thought to originate well outside the source of
continuum radiation (the gas is not as highly ionized at larger radii,
so UV resonance lines are optically thick enough for radiation
pressure to drive the wind).  Thus, although we see the hot continuum
source through the cooler wind, the wind can produce significant
emission lines (in addition to strong absorption) because the wind's
emitting area is much larger than the area of the continuum source
\citep{murray95}.  In tidal disruption events, by contrast, those
emitting areas are typically the same: the region where most line
emission occurs and the outer edge of the continuum source are both
typically at the electron scattering photosphere.  This, along with
the small true optical depth, implies that tidal disruption outflows
are expected to show little or no line emission.  The situation may be
more complicated if the outflow does not fully thermalize, as there
could be additional line emission inside the electron scattering
photosphere that we have not considered here.\footnote{If the outflow
  runs into relatively dense circumnuclear gas, additional line
  emission may be possible there as well.}  In addition to this
important difference in the presence of emission lines, note that BAL
QSOs are generally associated with higher-mass BHs ($\mbh \gtrsim
10^8\msun$), which cannot tidally disrupt solar-type stars outside the
horizon.

Another significant contaminant in optical searches for tidal
disruption events is supernovae that take place in the nuclei of
galaxies, because the luminosities and timescales are similar, and
because supernovae are initially blue like tidal disruption events.
Spectroscopy will help: supernovae typically show a forest of deep
optical absorption lines, quite unlike our predictions for tidal
disruption events.  Type IIn supernovae may still pose a particular
challenge: the supernova plows into surrounding circumstellar medium,
shock-heating the gas to high enough temperatures to suppress the
optical lines.  However, these supernovae begin to cool after several
months, becoming redder and producing optical emission lines at late
times.  Tidal disruption events, by contrast, become {\em hotter} with
time and are not expected to produce similar optical emission lines.

To quantify how much nuclear supernovae will `contaminate' searches
for tidal disruption events, we estimated the rate of supernovae
within the spatial resolution ($\sim 0.5$'') of ground-based optical
transient surveys (\S \ref{SNrates} and Fig. \ref{SNratesfig}): at
$z\sim 0.1$ the nuclear Type II rate is a ${\rm few} \times
10^{-3}\yr^{-1}$ per galaxy, of which $\sim$ 10\% are probably Type
IIn \citep{li10}; the Type Ia rate is a factor of $2-3$ smaller than
the Type II rate.  These rates of Type II and Type IIn supernovae are
$\gtrsim 2$ and 1 order(s) of magnitude greater than the tidal
disruption rate per galaxy inferred by \citet{donley} using ROSAT. The
ROSAT constraints, however, are largely on more massive black holes
(which are both more luminous and more likely to be prominent in the
X-rays) so it is possible that the tidal disruption rate is different
in the lower mass systems most likely to produce optically luminous
emission from super-Eddington outflows.  The high rate of nuclear
supernovae estimated in Figure \ref{SNratesfig} emphasizes the
importance of high-resolution follow-up imaging with HST or adaptive
optics, which can reduce the rate of nuclear supernovae `false
positives' in optical tidal disruption searches by a factor of
$\gtrsim 50$ (Fig. \ref{SNratesfig}).

Although the super-Eddington outflows produced during the tidal
disruption of stars are in principle readily detectable out to $z
\gtrsim 1$ \citep{sq09}, restricting candidate events to hosts with $z
\sim 0.1$ is likely a good strategy for minimizing interference from
supernovae: the rate of nuclear supernovae increases rapidly with
redshift, to $\sim 10^{-1}\yr^{-1}$ per galaxy at $z \sim 1$. This
increase is due to both the increase in star formation at high
redshift and the increasing fraction of a galaxy that lies within the
point-spread function of the observation.

The fiducial outflow model in \citet{sq09} ($f_{\rm out} = 0.1$; $f_v
= 1$; disruption rate per galaxy = $10^{-5} \, {\rm yr^{-1}}$)
predicts that $\sim 8$ tidal disruptions per year can be detected at
$z \lesssim 0.1$ for a survey like the Palomar Transient Factory.
Relatively nearby events have the additional advantages that they are
less expensive to follow up spectroscopically and it is easier to
characterize their host galaxies.  For these comparatively nearby
events, we predict that the detection probability in optical surveys
is relatively independent of $\mbh$ for $\mbh \lesssim {\rm few}\times
10^7\msun$, so it should help to restrict candidates to bulges less
massive than $\sim 10^{10}\msun$ (the weak dependence of the detection
probability on $\mbh$ is also important because tidal disruption
flares from lower mass BHs $\sim 10^6 M_\odot$ are the most likely to
be in thermal equilibrium and thus to have prominent optical continuum
emission in the first place).  Restricting follow-up to systems with
little ongoing star formation and/or old stellar populations would
help further minimize the number of nuclear supernovae.
Observationally, the star formation rate in galaxies, and hence the
supernova rate, is relatively bimodal, and thus some galaxies will
have nuclear supernova rates (particularly type II rates) smaller than
estimated in Figure \ref{SNratesfig}.  This is especially true for
more massive BHs, but $\sim 10 \%$ of lower massive systems ($\mbh
\lesssim 10^7 \msun$) are likely to be relatively passive as well
\citep[e.g.,][]{bernardi10}.  If these can be identified in advance
(via, e.g., prior Sloan Digital Sky Survey observations), they may
well be the most promising systems in which to follow up nuclear
transients; this selection would, however, decrease the predicted
detection rate to $\sim1\yr^{-1}$ at $z \lesssim 0.1$ for a survey
like PTF.

Given the recent plethora of luminous supernovae
\citep[e.g.,][]{quimby09}, it is unclear how well single-band
photometry alone will be able to distinguish tidal disruption events
from nuclear supernovae.  If this proves difficult, as we suspect is
likely, the spectroscopic predictions for tidal disruption events
presented here, and the color evolution predicted in \citet{sq09} (the
outflow photosphere becomes hotter with time---becoming bluer if the
observing band is close to the peak, or showing no color evolution if
the observing band is on the Rayleigh-Jeans tail), may prove
particularly useful for identifying and characterizing tidal
disruption flares.

\subsection{Optically-selected candidates}
The first two optically-selected tidal disruption candidates were
announced shortly after the submission of this paper, found in Stripe
82 of the Sloan Digital Sky Survey \citep{vanvelzen10}.  These
candidates have observed properties broadely consistent with our
predictions, with observed $g$-band luminosities of $\sim
10^{43}\erg\s^{-1}$ and no color evolution.  The optical data for
candidate ``TDE1'' is reasonably approximated by our model for the
super-Eddington outflow due to the disruption of a solar-type star by
a BH of mass $\sim 10^7\msun$ at $\rp\sim \rt$, with $f_{\rm out} \sim
0.1$ [eq. \ref{mdotout}] and $f_v \sim 0.1$ [eq. \ref{vwind}].  In
detail, the optical data for candidate ``TDE2'' is harder to
approximate with our simple model: although the luminosity can be
reproduced by a relatively large $f_{\rm out}$ or relatively small
$f_v$, the optical colors and only gentle fading of the event are less
consistent with our model.  These properties may be more consistent
with a model like that of \citet{loeb97} in which falling-back gas
settles into a steady hydrostatic atmosphere rather than becoming
unbound in a true outflow; this may be appropriate if the fallback
rate is never highly super-Eddington.  The large optical luminosity of
these two candidate events suggests that outflows may have relatively
low velocities, and is encouraging for the optical detection of future
tidal disruption candidates.

\section*{Acknowledgments}
We thank Josh Bloom, Kevin Bundy, Phil Chang, Weidong Li, Norm Murray,
Kristen Shapiro, and Nathan Smith for helpful conversations; we also
especially thank David Strubbe for help with C++ calculations.
Support for EQ was provided in part by the Miller Institute for Basic
Research in Science, University of California Berkeley and the David
and Lucile Packard Foundation.  Support for program number AR 12151
was provided by NASA through a grant from the Space Telescope Science
Institute, which is operated by the Association of Universities for
Research in Astronomy, Inc., under NASA contract NAS5-26555.

\bibliography{refs}

\label{lastpage}

\end{document}